\documentclass[sigconf,authorversion,nonacm]{acmart}
\usepackage{subfigure}
\usepackage{bm}
\usepackage{multirow}
\usepackage{comment}

\AtBeginDocument{%
  }

\acmConference[SBGames 2023]{22nd Brazilian Symposium on Games and Digital Entertainment}{November 06--09, 2023}{Rio Grande (RS), Brazil}
\acmBooktitle{22nd Brazilian Symposium on Games and Digital Entertainment (SBGames 2023), November 06--09, 2023, Rio Grande (RS), Brazil}
\begin{document}

\title{Revisiting Micro and Macro Expressions in Computer Graphics Characters}

%Vic: tentei colocar no titulo o seguinte:  - Comparisons between Different Levels of Realism and Different Years
%mas o tamanho ficou grande no header


\author{Rubens Montanha, Giovana Raupp, Vitoria Gonzalez, %\and Rafael Geiss, \and Márcio Pinho, 
Yanny Partichelli, André Bins, \\ Marcos Ferreira, Victor Araujo and Soraia Musse}
\affiliation{%
 \institution{Virtual Humans Lab - Graduate Program in Computer Science \\ Pontifical Catholic University of Rio Grande do Sul}
 \city{Porto Alegre}
 \country{Brazil}
 \\
Email: \{rubens.montanha,giovana.nascimento,v.gonzalez22,yanny.p,a.bins,\\marcos.nascimento,victor.araujo\}@edu.pucrs.br, soraia.musse@pucrs.br}

\renewcommand{\shortauthors}{Montanha et al.}

\begin{abstract}
  This paper presents the reproduction of two studies focused on the perception of micro and macro expressions of Virtual Humans (VHs) generated by Computer Graphics (CG), first described in 2014 and replicated in 2021. The 2014 study referred to a VH realistic, whereas, in 2021, it referred to a VH cartoon. In our work, we replicate the study by using a realistic CG character. Our main goals are to compare the perceptions of micro and macro expressions between levels of realism (2021 cartoon versus 2023 realistic) and between realistic characters in different periods (i.e., 2014 versus 2023). In one of our results, people more easily recognized micro expressions in realistic VHs than in a cartoon VH. In another result, we show that the participants' perception was similar for both micro and macro expressions in 2014 and 2023.
\end{abstract}

\begin{CCSXML}
<ccs2012>
   <concept>
       <concept_id>10010147.10010371.10010387.10010393</concept_id>
       <concept_desc>Computing methodologies~Perception</concept_desc>
       <concept_significance>500</concept_significance>
       </concept>
    <concept>
       <concept_id>10010147.10010371.10010387.10010393</concept_id>
       <concept_desc>Perception~Emotions and Faces</concept_desc>
       <concept_significance>500</concept_significance>
       </concept>
   <concept>
       <concept_id>10003120.10003121</concept_id>
       <concept_desc>Human-centered computing~Human computer interaction (HCI)</concept_desc>
       <concept_significance>300</concept_significance>
       </concept>
   <concept>
       <concept_id>10010147.10010371.10010352</concept_id>
       <concept_desc>Computing methodologies~Animation</concept_desc>
       <concept_significance>300</concept_significance>
       </concept>
 </ccs2012>
\end{CCSXML}

\ccsdesc[500]{Computing methodologies~Perception}
\ccsdesc[500]{Perception~Emotions and Faces}
\ccsdesc[300]{Human-centered computing~Human computer interaction (HCI)}
\ccsdesc[300]{Computing methodologies~Animation}

\keywords{Computer Graphics, Perception, Micro and Macro Expressions, Faces, Virtual Humans}


\maketitle

\section{Introduction}
\label{sec:introduction}
Over the years, the advance of Computer Graphics (CG) allowed the creation of more realistic virtual humans, making their facial and body animations more similar to real human beings.~\footnote{Draft version made for arXiv: https://arxiv.org/} %expressions similar to reality. 
Some studies in facial expression propose universal facial expressions of emotion \cite{ekman1974nonverbal, darwin1872expression}. However, these works also say that these facial expressions are inherited and can be different due to personal, social, or even cultural conditions suppressing their real emotions. Following Ekman and Friesen \cite{ekman1969nonverbal}, some methods in literature present facial emotions using macro facial expressions and micro facial expressions. The macro are facial expressions that are obvious to be perceived and during between 0.5 and 4 seconds. This kind of facial expression typically matches the expression with the context. Oppositely, micro expressions are often misinterpreted or not perceived during less than 500 milliseconds, being expressed unconsciously. % in terms of realistic Virtual Humans (VHs). 
According to Ekman \cite{ekman2005argument}, the macro and micro expressions can be classified into six emotions: anger, fear, enjoyment, happiness, surprise, and disgust. Macro expressions are shorter than other body expressions, like feet and legs\cite{ekman1969nonverbal}. %Micro expressions are shorter than macro expressions, which can be part of a macro expression and appear in milliseconds. %Being shorter than others, micro expressions  can be more universally recognized. 
%SO: acima quem falou essa coisa de legs.. colocar ref para isso
%RU: Pronto. Referencia colocada.
Still, according to Ekman, facial expressions are sets of Action Units (AU), which consist of specific parts of faces, forming a system called Facial Action Coding system (FACs)~\cite{ekman1978facial, friesen1983emfacs}. In terms of VHs, the representation of human emotions aims to create realistic behaviors similar to real human beings, making the audience (people) feel comfortable~\cite{mori2012uncanny, araujo2021perceived, katsyri2015review}. To evaluate the human perception of emotions, the FACs system is presented in scientific studies~\cite{ekman1978facial, friesen1983emfacs}. In addition, it is also presented in various media, such as movies~\footnote{https://www.fxguide.com/fxfeatured/making-thanos-face-the-avengers/}.

Focusing on VHs, Zell et al. \cite{zell2019perception} present the importance of perception and how people perceive virtual characters. In their work, the authors explain a general perception model where a person sees an object, which is used as a visual input. This visual input is processed in the brain following bottom-up and top-down processes. This process of perceiving and recognizing these objects involves transforming low-level sensory information into high-level information, the bottom-up process, and the person's cognitive states, including his/her personality and motivation. These cognitive states can be associated with schemes and include the person's life experiences. At the end of this process, the person recognizes the object. This general model can also be applied in the perception of virtual characters \cite{zell2019perception}. Understanding how the 3D depth and shapes are perceived in a 2D representation is important for the design of new virtual humans using pre-existing schemes. Based on the character's appearance, people can add attributes and personalities provided by their judgments.
%SO: a frase acima parece incompleta
%RU: Mudei a última frase e adicionei uma para justificar essa questão de percepção.
Zell et al.'s work reinforces the significance of perceiving and interpreting other people's emotions for successful social interaction. Consequently, when a character accurately expresses emotions, individuals are likelier to find it engaging.

Specifically about micro and macro expressions in VHs, Queiroz et al. \cite{queiroz2014investigating} studied, in 2014, the users' perception of hidden emotions in virtual faces, investigating if the way people perceive the facial expressions with realistic virtual characters was similar to real people. Another hypothesis studied by the authors was if facial micro expressions could, in some situations, show a second emotion, even when the person or virtual human is expressing a neutral facial expression. Given the potential advancements in technology since 2014, is it likely that the results about facial emotion recognition would differ when applied to a realistic VH generated using more recent technological capabilities? Following the same methodology, Andreotti and collaborators \cite{andreotti2021perception} studied the perception of micro and macro facial expressions using a cartoon character in 2021. In their work, the authors also showed the macro and micro expressions of positive and negative emotions, and they evaluated how the character is charismatic and comfortable, according to the users' perception. So, another question aims to discuss if varying levels of realism, as explored in Andreotti et al.'s work and ours, would present some disparity in emotion recognition. %n between cartoon-style and realistic VHs persist when evaluating a realistic VH generated using more advanced technologies?

To answer the questions raised in the previous paragraph, this work presents a study of how people perceive realistic VH facial expressions in 2023. Following the same methodology introduced by Queiroz et al. \cite{queiroz2014investigating} and used by Andreotti et al.\cite{andreotti2021perception}, we elaborated two hypotheses to be answered following this work: 

\begin{itemize}
    \item $H0_1$ - There is no significant difference in the perception of micro and macro facial expressions between realistic and cartoon virtual humans; and
    \item $H0_2$ - There is no significant difference in the perception of micro and macro facial expressions between realistic virtual humans in 2014 and those in 2023.
\end{itemize}

% and  %and \textit{3)} ''Are the perception of facial expressions of female CG characters similar to male characters?" 
To try to answer these hypotheses, we recreated the two experiments of %proposed in 
Queiroz et al. \cite{queiroz2014investigating} and Andreotti et al.~\cite{andreotti2021perception}. Our main goal in this work is to observe changes in realistic virtual human expressions between 2014 and 2023 and analyze the similarities in how people perceive cartoon characters and realistic virtual humans. 

%Vic: pensando ainda se usamos o personagem masculino nesse paper.. ta meio fora do contexto

The rest of this work is organized as follows. Section \ref{sec:related_work} presents some related work in the literature, while Section \ref{sec:methodology} explains the methodology used in this work. Results are presented and explained in Section \ref{sec:results}, whereas Section \ref{sec:discussion} discusses the results found in this work. Finally, Section \ref{sec:final_considearions} closes this work and presents our final considerations and future works.

% This work presents a study of how people perceive realistic virtual humans face expressions in 2023, following the same methodology introduced by Queiroz et al. \cite{queiroz2014investigating} and used by Andreotti et al.\cite{andreotti2021perception}. Our main goal is to investigate the similarity of realistic characters expressions in 2014 and in 2023, and the similarity between realistic and cartoon facial expressions. As well, we analysed the similarity between the male and female facial expressions in virtual characters.

% \begin{itemize}
%     \item H$0_1$ The perception of micro and macro facial expressions of realistic virtual humans is similar to cartoon characters.
%     \item H$0_2$ The perception of micro and macro facial expressions of virtual humans in 2014 is similar in 2023
%     \item H$0_3$ The perception of facial expressions of female CG characters is similar to male characters
% \end{itemize}

\section{Related Work}
\label{sec:related_work}

This section presents studies on emotion perception regarding VHs and human beings. Humans have the ability to recognize and categorize human behavioral movements even from minimal information~\cite{johansson1973visual}. Emotions are an integral part of human behaviors, and several studies have focused on understanding emotional expressions \cite{friesen1983emfacs, ekman1978facial, ekman2005argument, bassili1978facial}. Concerning micro expressions, the pioneering work of Gottschalk et al.~\cite{gottschalk1966micromomentary}, who analyzed psychotherapy cinematographic films and identified micro expressions as micro momentary expressions. Their study aimed to uncover nonverbal communication between patients and therapists. In their investigations, it was observed that the expression on the patient's face would occasionally undergo drastic changes within three to five frames, equivalent to a period of 1/8 to 1/5 of a second, transitioning from a smile to a frown and back to a smile. Subsequently, Paul Ekman formally coined the term "micro expressions"~\cite{ekman1969nonverbal}. Presently, there is a significant research interest in this field, particularly in its application to lie detection~\cite{matsumoto2011evidence}.

Our research draws inspiration from the studies by Queiroz et al. \cite{queiroz2014investigating} and Andreotti et al.~\cite{andreotti2021perception}. The first one was focused on three psychological investigations that explored the perceptions of micro and macro expressions. One of these investigations, proposed by Bornemann et al.~\cite{bornemann2012can}, involved the brief presentation of micro expressions lasting between 10 and 20 milliseconds to ensure that individuals did not consciously perceive them. Another investigation by Shen et al.~\cite{shen2012effects} 
utilized two methodologies, BART (Brief Affect Recognition Test) and METT (Micro Expression Training Tool), based on Ekman's research. In the BART condition, participants were shown the six universal expressions following a fixed point, while in the METT paradigm, the universal expressions were presented between two sequences of neutral faces. That study aimed to determine the upper limit of time required for perceiving micro expressions and concluded that the accuracy of participants' responses began to stabilize at 160 milliseconds. Lastly, the study of Li et al.~\cite{li2008neural} involved presenting the expression of a surprise following 30 milliseconds of either happiness or fear and then asking observers to determine whether they perceived the preceding expressions as positive or negative. Andreotti et al.~\cite{andreotti2021perception} used these three psychological investigations based on the technique proposed in Queiroz et al. to study micro and macro expressions at a different level of realism. The authors used a cartoon VH and compared their results with the work of Queiroz et al. (realistic VH).
%SO: se tudo acima é sobre as investigações no paper da queiroz, a gente não fala nada do Andreotti?
%RU: Coloquei uma ligação entre o trabalho da Queiroz e do Andreotti na última linha.

Emotions are present in studies involving VHs. For example, Melgaré et al. \cite{Melgare22, Melgare2019} presented a method for preprocessing images of real faces from a specific group of human beings. They created an average face that could be manipulated to produce exaggerated or smoothed representations of the group, referred to as facial style. The perception of emotions is also closely related to VHs \cite{ennis2013emotion, andreotti2021perception, queiroz2014investigating, hyde2014assessing, hyde2016evaluating, ochs201518, krumhuber2016perceptual, bailey2017gender, zibrek2013evaluating}. For example, the study of Ennis et al. \cite{ennis2013emotion} conducted a perceptual experiment using synchronized full-body and facial motion-capture data. The findings indicate that individuals can recognize emotions from either body or facial motion alone. The authors evaluated four macro expressions (full emotions): Anger, Fear, Happiness, and Sadness. Regarding facial emotion, Happiness was the only one with a percentage of correct answers above 70\% of the participants, and Fear had the lowest percentage (below 40\%). The study of Hyde et al.~\cite{hyde2014assessing} investigated the impact of altering auditory and facial expressiveness levels on emotion recognition accuracy, perceived emotional intensity, and naturalness ratings. In the proposed experiment, participants evaluated animations of a character whose facial motion matched a tracked actress's. The study found that higher auditory expressiveness positively influenced emotion recognition accuracy and emotional intensity ratings.

%Vic: aqui falar sobre queiroz e andreotti pra fzer o link com a prox section
In our work, we follow Queiroz et al. \cite{queiroz2014investigating} and Andreotti et al. \cite{andreotti2021perception} methodologies to compare the similarities between the Andreotti et al. and our results ($H0_1$), and Queiroz et al. and our results ($H0_2$). Queiroz et al. work analyses the perception of realistic virtual human facial expressions in 2014, comparing results of how the participants perceive and recognize macro and micro expressions compared with BART and METT. Following the same idea and methodology, Andreotti et al., in 2021, analyzed the perception and recognition of a cartoon character's macro and micro expressions by the participants. In their work, the authors also compared the results between the cartoon VH and the realistic VH (the results of Queiroz et al. work). The following section presents these methodologies and explains how our study was conducted.

\section{Methodology}
\label{sec:methodology}

This section aims to present the methodology used in this work, which was based on those mentioned in Section~\ref{sec:related_work} (studies of Andreotti et al.~\cite{andreotti2021perception} and Queiroz et al.~\cite{queiroz2014investigating}). We used this methodology to answer $H0_1$ and $H0_2$. The authors performed two sets of experiments in the methodology proposed by Queiroz et al. For each one, they prepared a survey with questions about the universal facial expressions in some short videos presented during the question. The first experiment was performed via the Internet, while the second was done in person, with supervision over the participants. In addition, the authors used a realistic VH. The methodology used by Andreotti et al. performed the survey in the same way as proposed by Queiroz et al., but only Experiment 1 was executed due to pandemic reasons. In that case, the authors used a cartoon VH. 

As done in Andreotti et al. work, we prepared an online survey~\footnote{Project Estudos e Avaliações da Percepção Humana em Personagens e Multidões Virtuais, number 46571721.6.0000.5336, approved by the Ethics Committee of Pontifical Catholic University of Rio Grande do Sul.}, which was applied using the Qualtrics survey tool. The form and the animations were made based on Queiroz et al. and Andreotti et al. studies. Furthermore, all questions used in our work were based on these two studies. The form was divided into five parts:
\begin{enumerate}
    \item Consent form – The purpose of the research was explained to the participant, and agreement with the terms explained was required.
    \item Demography questions – Personal information about the participant, such as gender, education, age (average), and familiarity with computer graphics, was collected.
    \item Control questions – Six videos with visual macro expressions were first shown to determine if the participant accurately perceived them. Then, questions were asked about each presented expression.
    \item Experiment Part 1 – It comprised six videos of micro expressions and questions about each one.
    \item Experiment Part 2 – It was composed of 20 videos of micro expressions followed by a macro expression and questions about each one.
\end{enumerate}

\begin{figure*}
    \centering
        \subfigure[Anger]{
            \includegraphics[width=0.15\textwidth]{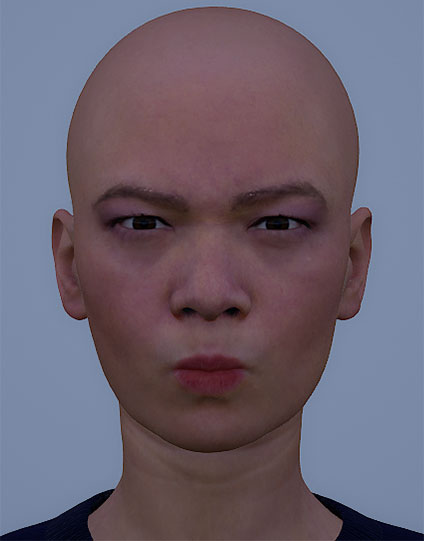}
            \label{fig:anger}
        }
        \subfigure[Disgust]{
            \includegraphics[width=0.15\textwidth]{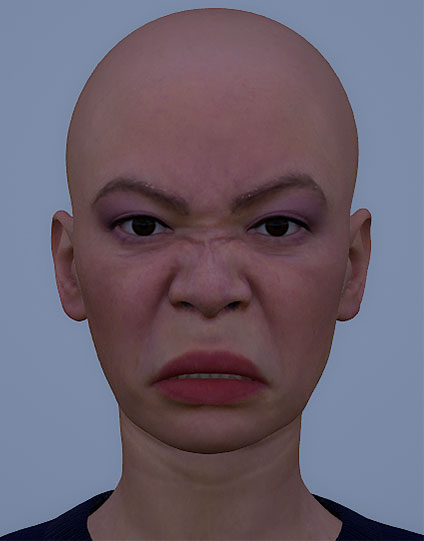}
            \label{fig:disgust}
        }
        \subfigure[Fear]{
            \includegraphics[width=0.15\textwidth]{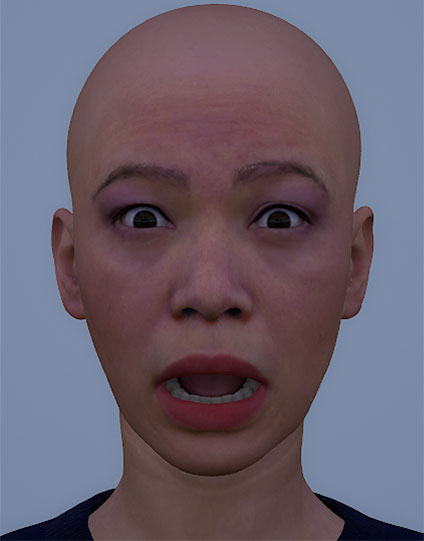}
            \label{fig:fear}
        }
        \subfigure[Happy]{
            \includegraphics[width=0.15\textwidth]{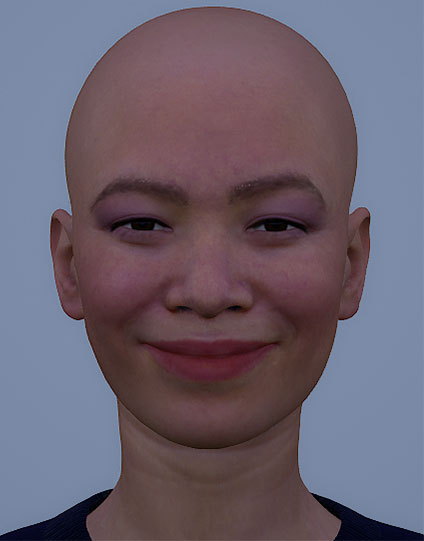}
            \label{fig:happy}
        }
        \subfigure[Sad]{
            \includegraphics[width=0.15\textwidth]{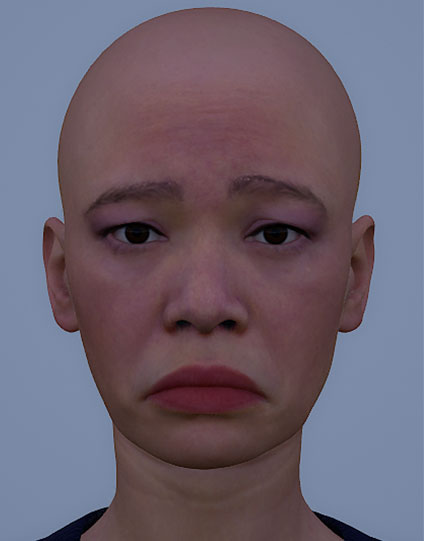}
            \label{fig:sad}
        }
        \subfigure[Surprise]{
            \includegraphics[width=0.15\textwidth]{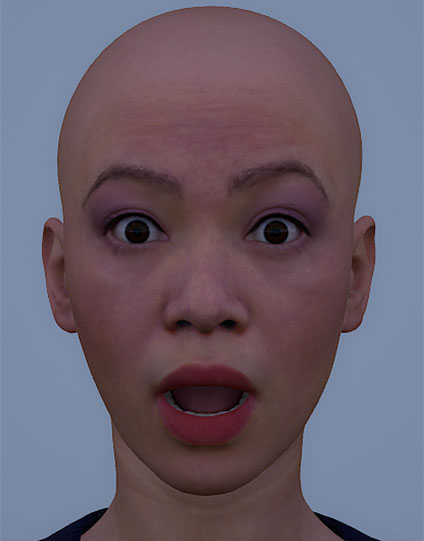}
            \label{fig:surprise}
        }
    \caption{All expressions represented by the realistic character presented in our work.}
    \label{fig:our_emotions}
\end{figure*}

To prepare the survey, we model the character used in this study using the MetaHuman Creator~\footnote{https://www.unrealengine.com/en-US/metahuman}, a framework to create and animate highly realistic digital human characters. Figure \ref{fig:our_emotions} presents this virtual human with six universal emotions. Metahuman provides a range of VHs, and as in the two baseline studies, we used a female, hairless VH model to avoid gender and hair bias. To create the Ekman-based emotions~\cite{ekman1978facial, ekman1969nonverbal, ekman1974nonverbal, ekman2005argument}, we exported the VH model to the Unreal engine via the Quixel Bridge software. With this, we use the VH facial rig provided by Unreal and manually adjust the sliders' values for each part of the VH face, %. Values were adjusted 
based on images of emotions and AUs trained in the Ekman style~\footnote{https://imotions.com/blog/learning/research-fundamentals/facial-action-coding-system/ \label{fot:imotions}}. 

\begin{figure*}
    \centering
        \subfigure[Anger]{
            \includegraphics[width=0.15\textwidth]{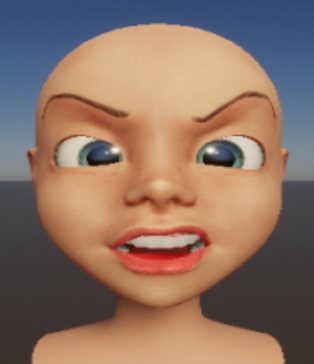}
            \label{fig:anger_andreotti}
        }
        \subfigure[Disgust]{
            \includegraphics[width=0.15\textwidth]{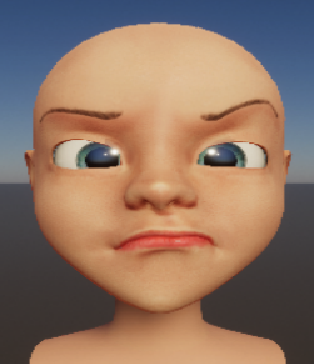}
            \label{fig:disgust_andreotti}
        }
        \subfigure[Fear]{
            \includegraphics[width=0.15\textwidth]{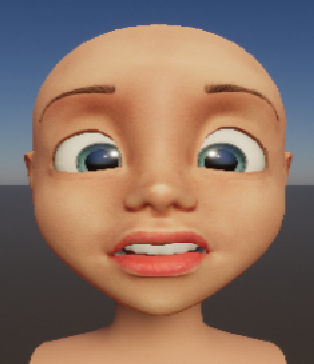}
            \label{fig:fear_andreotti}
        }
        \subfigure[Happy]{
            \includegraphics[width=0.15\textwidth]{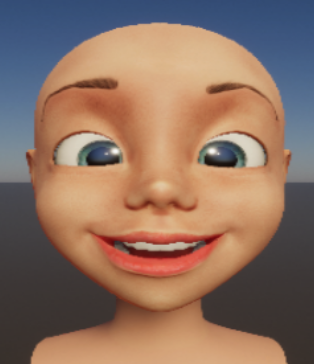}
            \label{fig:happy_andreotti}
        }
        \subfigure[Sad]{
            \includegraphics[width=0.15\textwidth]{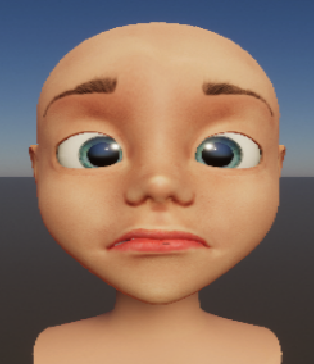}
            \label{fig:sad_andreotti}
        }
        \subfigure[Surprise]{
            \includegraphics[width=0.15\textwidth]{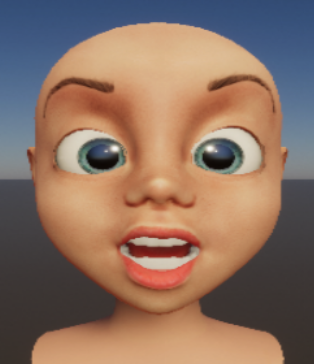}
            \label{fig:surprise_andreotti}
        }
    \caption{All expressions represented by the cartoon character presented in Andreotti et al. \cite{andreotti2021perception} work.}
    \label{fig:andreotti_emotions}
\end{figure*}

\begin{figure*}
    \centering
        \subfigure[Anger]{
            \includegraphics[width=0.15\textwidth]{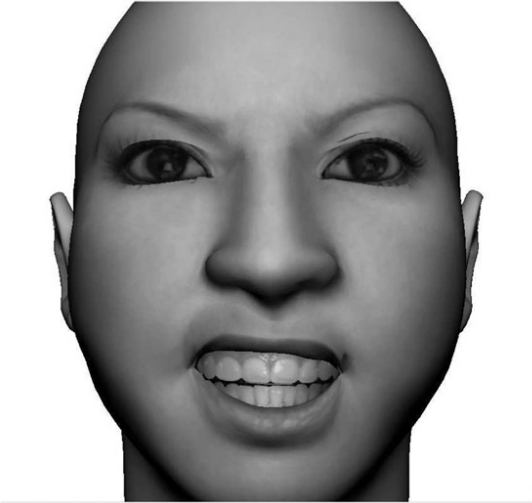}
            \label{fig:anger_queiroz}
        }
        \subfigure[Disgust]{
            \includegraphics[width=0.15\textwidth]{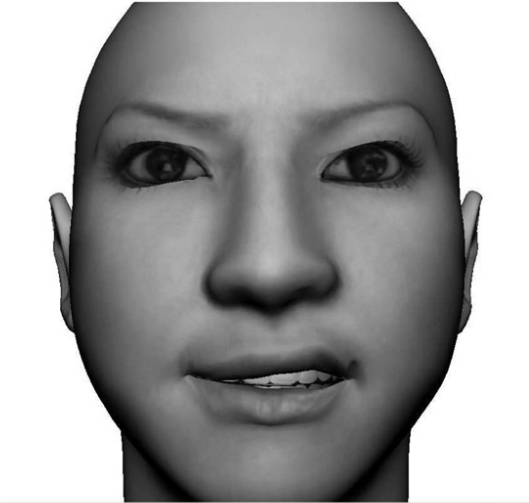}
            \label{fig:disgust_queiroz}
        }
        \subfigure[Fear]{
            \includegraphics[width=0.15\textwidth]{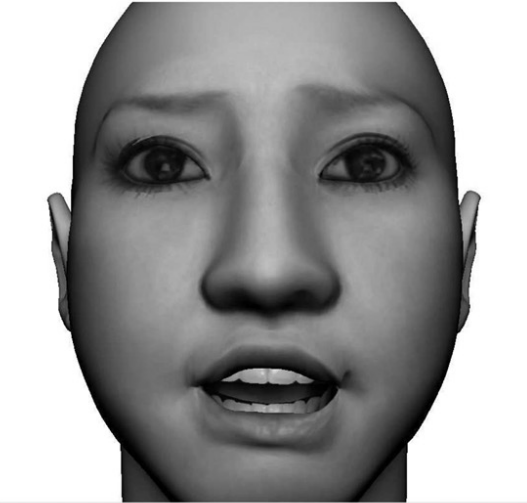}
            \label{fig:fear_queiroz}
        }
        \subfigure[Happy]{
            \includegraphics[width=0.15\textwidth]{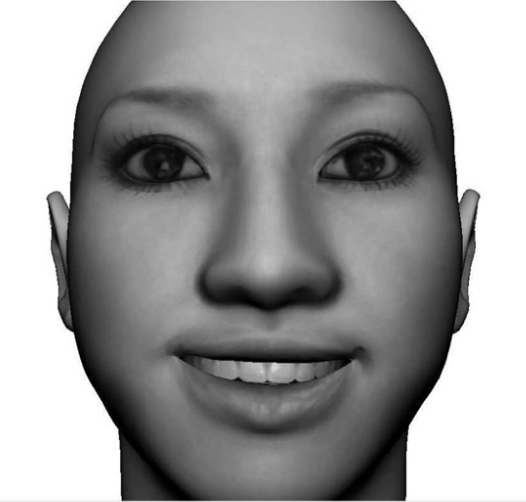}
            \label{fig:happy_queiroz}
        }
        \subfigure[Sad]{
            \includegraphics[width=0.15\textwidth]{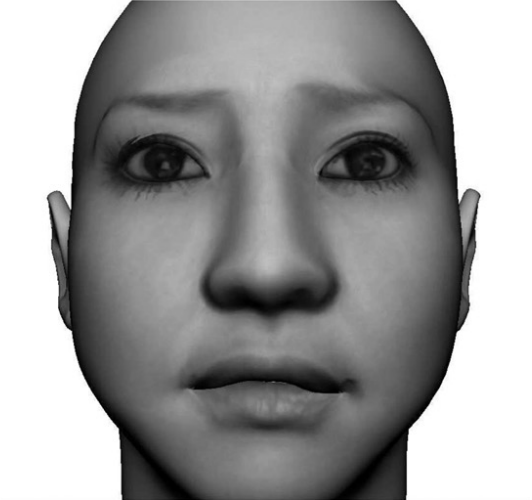}
            \label{fig:sad_queiroz}
        }
        \subfigure[Surprise]{
            \includegraphics[width=0.15\textwidth]{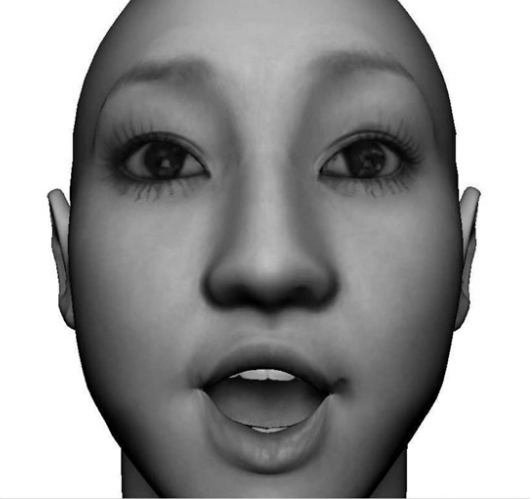}
            \label{fig:surprise_queiroz}
        }
    \caption{All expressions represented by the realistic character presented in Queiroz et al. \cite{queiroz2014investigating} work in 2014.}
    \label{fig:queiroz_emotions}
\end{figure*}

In Queiroz et al. methodology, the authors used a facial animation model based on blendshape interpolation. To create the animations, they used a facial animation control tool proposed by Queiroz et al. \cite{DBLP:journals/cie/QueirozCM09} and the FaceGen software, which generates 3D faces. Figure \ref{fig:queiroz_emotions} presents the realistic character and expressions used in Queiroz et al. work. Meanwhile, the methodology proposed by Andreotti et al. used a generic cartoon character bought on the Internet, which supports blendshapes, as suggested in Queiroz et al. work. The authors used the blendshapes to perform the animations of facial expressions and Unity Engine to generate the face expression videos, also based on Ekman style~\footref{fot:imotions}. Figure \ref{fig:andreotti_emotions} shows the cartoon character and expressions used in Andreotti et al.

\subsection{Control Questions}
\label{sec:contron_questions}

After the consent form and demography questions, the six emotions presented in Figure~\ref{fig:our_emotions} are shown to the participant. Then, six macro expressions videos are presented, followed by the control questions "What was the emotion presented in the video?" and "How many times did you need to see the video to be able to answer the question?". The options available to answer the first question are the six universal emotions (\cite{ekman1974nonverbal}), while the participant needs to answer the second with a number.

This step is important to evaluate if the participant can perceive the main expressions, as presented in Andreotti et al. \cite{andreotti2021perception} methodology. %This part is followed by the two experiments explained previously in Section~\ref{sec:methodology}. 
Table~\ref{tab:videos_experiments} presents the order of the videos presented to the users. As followed by Andreotti et al., we recreated the six videos from Part 1 and 10 from Part 2 proposed by Queiroz et al. \cite{queiroz2014investigating}. We also recreated more ten videos produced in Andreotti et al. work. Next sections detail the performed experiments.

\subsection{Experiment Part 1 }
\label{sec:experiment_1}

This part of the experiment presents to the participant a single micro expression video for each universal emotion following Table~\ref{tab:videos_experiments}. These videos presented the micro expression with a duration of 100ms between two sequences of 2 seconds with a neutral face. After each video, the question "What emotion is present in this video?" was asked with the possible answers: Anger, Disgust, Fear, Happy, Sad, Surprise, I don’t know, and No emotion.

\subsection{Experiment Part 2}
\label{sec:experiment_2}

The second part of the experiment evaluates the micro expression followed by the macro expression. In this step, it was used a total of 20 videos; 10 recreated using the evaluations proposed by Queiroz et al. \cite{queiroz2014investigating} and the other 10 proposed by Andreotti et al. \cite{andreotti2021perception}. The videos were composed of 100ms of a micro expression followed by 510ms of a macro expression, both jointed between two sequences of 300ms with a neutral expression. The second column of Table~\ref{tab:videos_experiments} presents the order of the videos and emotions used in this experiment.

After each video, three questions were asked to the participants. The questions were "Which main emotion was presented?", followed by "Do you think she’s feeling something else? Which emotion?" and "How many times did you have to watch the video to answer this question?". The possible answers to the first two questions are Anger, Disgust, Fear, Happy, Sad, Surprise, I don’t know, and No emotion, while the last question is a number.

\begin{table}[ht]
    \centering
    \caption{Order of the videos and emotions used in the experiments and in our work. The emotions in bold are the videos added by Andreotti et al. \cite{andreotti2021perception} in their methodology, while the other emotions are videos proposed in Queiroz et al. \cite{queiroz2014investigating} work.}
    \begin{tabular}{|c|c|c|}
        \hline
         \textbf{Experiment Part 1} & \multicolumn{2}{c|}{\textbf{Experiment Part 2}}  \\
        \hline
         \textbf{Micro Expression} & \textbf{Macro Expression} & \textbf{Micro Expression} \\
        \hline
         Anger & Fear & Anger \\
        \hline
         Disgust & \textbf{Happiness} & \textbf{Surprise} \\
        \hline
         Fear & \textbf{Fear} & \textbf{Surprise} \\
        \hline
         Happiness & \textbf{Surprise} & \textbf{Disgust} \\
        \hline
         Sadness & \textbf{Anger} & \textbf{Surprise} \\
        \hline
         Surprise & \textbf{Disgust} & \textbf{Fear} \\
        \hline
         & Happiness & Sadness \\
         \hline
         & Fear & Disgust \\
         \hline
         & Surprise & Happiness \\
         \hline
         & \textbf{Fear} & \textbf{Disgust} \\
         \hline
         & Sadness & Anger \\
         \hline
         & Anger & Happiness \\
         \hline
         & Surprise & Fear \\
         \hline
         & Surprise & Sadness \\
         \hline
         & \textbf{Disgust} & \textbf{Surprise} \\
         \hline
         & \textbf{Sadness} & \textbf{Disgust} \\
         \hline
         & Sadness & Happy \\
         \hline
         & Happiness & Anger \\
         \hline
         & \textbf{Fear} & \textbf{Happiness} \\
         \hline
         & \textbf{Disgust} & \textbf{Sadness} \\
        \hline
    \end{tabular}
    \label{tab:videos_experiments}
\end{table}

% Following the Queiroz et al. \cite{queiroz2014investigating} methodology, we performed one experiment. This experiment was made in a survey with questions about facial expressions in some short videos. The survey were applied online using the Qualtrics survey tool.

\section{Results}
\label{sec:results}

This section aims to present the results referring to the questionnaire presented in Section~\ref{sec:methodology}. Our questionnaire was answered by 30 volunteers. %where 56.66\% of them were women and 43.33\% were men
Table~\ref{tab:demographicPercentages} presents the demographic profile percentage results. %\red{Regarding the age of the participants, 46.67\% were under 36 years of age, while 53.33\% were 36 years of age or older. Regarding the familiarity with CG, 71.43\% were familiar with CG and 28.57\% were not. In relation to the level of education, 43.33\% of the participants had Complete High School (CHS), and 56.67\% had Complete Higher Education (CHE).} 
To try to avoid possible biases in comparisons between groups of different people (our work versus the work of Queiroz et al.~\cite{queiroz2014investigating} versus the work of Andreotti et al.~\cite{andreotti2021perception}), we tried to recruit volunteers in the same social networks and the same country, so people are culturally comparable. Demographic data from both previous studies are presented to conduct a visual comparison. In the work of Andreotti et al., the experiments were answered by 81 participants, with 53.42\% identifying as women and 46.57\% as men. Regarding age distribution, 67.12\% of participants were below 30 years old, while 32.87\% were aged 30 or above. Concerning familiarity with CG, 54.79\% of participants reported being familiar, whereas 45.20\% indicated unfamiliarity. Furthermore, Queiroz et al. used a different data collection methodology, which involved gathering responses from 84 participants over a five day period. All these participants were individuals from academia, comprising 72 males and 12 females, and they had an average age of 27.57 with a standard deviation of 9.15. %As for Experiment 2, collected the responses over a 3 day period with 10 participants, also academics. This group included seven males and three females, with an average age of 26.4 and a standard deviation of 7.06. Notably, two volunteers took part in both sets of experiments.
  
\begin{table*}[ht]
\centering

\caption{Demographic percentages of voluntary participants in our work.}

  \label{tab:demographicPercentages}
  %\begin{adjustbox}{max width=0.98\linewidth}
\begin{tabular}{|c|cc|cc|cc|cc|}
\hline
\textbf{}   & \multicolumn{2}{c|}{\textbf{Gender}}             & \multicolumn{2}{c|}{\textbf{Age}}                                & \multicolumn{2}{c|}{\textbf{Education}}          & \multicolumn{2}{c|}{\textbf{Familiarity}}        \\ \hline
\textbf{Total} & \multicolumn{1}{c|}{\textbf{Female}}  & \textbf{Male}    & \multicolumn{1}{c|}{\textbf{\textless{}36}} & \textbf{\textgreater{}=36} & \multicolumn{1}{c|}{\textbf{CHS}}     & \textbf{CHE}     & \multicolumn{1}{c|}{\textbf{Yes}}     & \textbf{No}      \\ \hline
\textbf{30}    & \multicolumn{1}{c|}{\textbf{56.67\%}} & \textbf{43.33\%} & \multicolumn{1}{c|}{\textbf{46.67\%}}       & \textbf{53.33\%}           & \multicolumn{1}{c|}{\textbf{43.33\%}} & \textbf{56.67\%} & \multicolumn{1}{c|}{\textbf{71.43\%}} & \textbf{28.57\%} \\ \hline

\end{tabular}
  %\end{adjustbox}
\end{table*}
We used a non-parametric paired \textit{T}-test (5\% significance level) to test our hypotheses. This section is distributed as follows: Section~\ref{sec:microIso} presents the results referring to micro and macro expressions (respectively, Experiment 1 and Control Questions) presented separately, while Section~\ref{sec:microFllowedMacro} presents the results referring to micro expressions presented before macro expressions (Experiment 2).

%\subsection{Unrealistic versus Realistic}
%\label{sec:unr_vs_real}

\subsection{Isolated Macro and Micro Expressions}
\label{sec:microIso}

%Vic: coloquei abaixo pra falar das isoladas nessa subsection
First of all, concerning the control questions (as shown in Table~\ref{tab:confusionMatrixEmotions_femaleParticipants_macro}), 82.77\% of the participants recognized the macro expressions. In the work by Andreotti et al., ~\cite{andreotti2021perception}, people had 80\% of correct answers, whereas Queiroz et al.~\cite{queiroz2014investigating} had no control questions.

\begin{table*}[ht]
  \centering
  \caption{
Confusion matrix with the percentages of all response options (Happiness, Sadness, Surprise, Fear, Anger, Disgust) of macro isolated animations. The participants' responses are presented along the columns' extensions following the top line. (Control Questions presented in Section~\ref{sec:contron_questions}).}
  \label{tab:confusionMatrixEmotions_femaleParticipants_macro}
  %\begin{adjustbox}{max width=0.98\linewidth}
  \begin{tabular}{|c|c|c|c|c|c|c|c|c|}
   \hline
   \textbf{Macro Isolated} & \textbf{Happiness} & \textbf{Sadness}  & \textbf{Anger} & \textbf{Surprise} & \textbf{Fear} & \textbf{Disgust}& \textbf{No Emotion}& \textbf{I Don't Know}\\
    \hline
    \textbf{Happiness} & \textbf{90.00\%} & 3.33\% & 3.33\% & 0.00\% & 3.33\% & 0.00\% & 0.00\% & 0.00\%\\ 
     \hline
      \textbf{Sadness} & 6.67\% & \textbf{80.00\%} & 0.00\% & 10.00\% & 3.33\% & 0.00\% & 0.00\% & 0.00\%\\ 
      \hline
    \textbf{Anger} & 00.00\% & 3.33\% & \textbf{83.33\%} & 6.67\% & 3.33\% & 3.33\% & 0.00\% & 0.00\%\\ 
     \hline
      \textbf{Surprise} & 3.33\% & 3.33\% & 6.67\% & \textbf{83.33\%}  & 3.33\% & 0.00\% & 0.00\% & 0.00\%\\
      \hline
      \textbf{Fear} & 0.00\% & 3.33\% & 0.00\% & 16.67\% & \textbf{80.00\%} & 0.00\% & 0.00\% & 0.00\%\\
     \hline
      \textbf{Disgust} & 3.33\% & 0.00\% & 10.00\% & 3.33\% & 3.33\% & \textbf{80.00\%} & 0.00\% & 0.00\%\\
    \hline

%    \bottomrule
  \end{tabular}
  %\end{adjustbox}
\end{table*}

In the videos containing only isolated micro expressions (Experiment Part 1), as we can see in Table~\ref{tab:confusionMatrixEmotions_maleParticipants_macro}, Fear's micro expression was the only one that had percentages of correct answers below 50\%. In the work by Andreotti et al. (as shown in Table~\ref{tab:confusionMatrixEmotions_unr_vs_rea_macro}), Sadness, Fear, Disgust, and Happiness had percentages below 50\%. While in the work by Queiroz et al., the result was similar to our work, that is, Fear was the emotion that had the lowest percentage and the only one below 50\%. In addition, comparing only the percentage of correct answers between the three studies, as highlighted in bold in Table~\ref{tab:confusionMatrixEmotions_unr_vs_rea_macro}, the work by Queiroz et al. did not have higher percentages of correct answers when the videos were about micro expressions of Sadness and Disgust, which had higher percentages of correct answers linked to our work. 
%SO: revisar a frase acima... "did not only"...???
%Vic: tirei o only
While the work by Andreotti et al. had the lowest percentages for all emotions recognition.

Regarding $H0_1$, we found a significant result (Stats=$2.44$ and \textit{p}=$.03$) in the comparison between the average percentages of micro expression recognition in our work (average of 71.11\%) versus the work of Andreotti et al.~\cite{andreotti2021perception} (AVG=38.05\%). With that, \textbf{refuting $H0_1$, we can say that people more correctly recognized the micro expression of the realistic VH of our work than of the cartoon VH of the work of Andreotti et al.} Concerning $H0_2$, we found no significant results. Even though the average percentage of micro expression recognition in our work was lower than in work by Queiroz et al.~\cite{queiroz2014investigating}(AVG=78.17\%). Therefore, \textbf{we cannot refute $H0_2$, that is, both realistic VH from 2014 (work by Queiroz et al.) and realistic VH from 2023 (our work) had micro expressions correctly recognized similarly.} We also performed an analysis comparing the average percentages between micro and macro expressions of our work. However, we found no significant result. In this case, \textbf{we can say that people correctly recognized both isolated micro and macro expressions.}

\begin{table*}[ht]
  \centering
  \caption{
Confusion matrix with the percentages of all participants' responses according to the emotional options (Happiness, Sadness, Surprise, Fear, Anger, Disgust) of micro isolated animations (Experiment Part 1).}
  \label{tab:confusionMatrixEmotions_maleParticipants_macro}
  %\begin{adjustbox}{max width=0.98\linewidth}
  \begin{tabular}{|c|c|c|c|c|c|c|c|c|}
   \hline
   \textbf{Micro Isolated} & \textbf{Happiness} & \textbf{Sadness}  & \textbf{Anger} & \textbf{Surprise} & \textbf{Fear} & \textbf{Disgust}& \textbf{No Emotion}& \textbf{I Don't Know}\\
    \hline
    \textbf{Happiness} & \textbf{83.33\%} & 3.33\% & 0.00\% & 6.67\% & 6.67\% & 0.00\% & 0.00\% & 0.00\%\\ 
     \hline
      \textbf{Sadness} & 3.33\% & \textbf{80.00\%} & 0.00\% & 6.67\% & 6.67\% & 0.00\% & 0.00\% & 0.00\%\\ 
      \hline
    \textbf{Anger} & 0.00\% & 10.00\% & \textbf{50.00\%} & 10.00\% & 16.67\% & 13.33\% & 0.00\% & 0.00\%\\ 
     \hline
      \textbf{Surprise} & 0.00\% & 0.00\% & 3.33\% & \textbf{90.00\%}  & 3.33\% & 3.33\% & 0.00\% & 0.00\%\\
      \hline
      \textbf{Fear} & 0.00\% & 6.67\% & 0.00\% & 53.33\% & \textbf{36.67\%} & 3.33\% & 0.00\% & 0.00\%\\
     \hline
      \textbf{Disgust} & 0.00\% & 3.33\% & 10.00\% & 6.67\% & 3.33\% & \textbf{86.67\%} & 0.00\% & 0.00\%\\
      \hline
%    \bottomrule
  \end{tabular}
  %\end{adjustbox}
\end{table*}

\subsection{Micro expression followed by Macro expression}
\label{sec:microFllowedMacro}

Regarding videos that had micro expressions followed by macro expressions (Experiment Part 2), Table~\ref{tab:confusionMatrixEmotions_femaleParticipants_micro} shows that the Sadness micro expression had the highest percentage of correct answers compared to the other micro expressions, being the only one that had a percentage above 40\%. Compared with the other two studies, Table~\ref{tab:confusionMatrixEmotions_unr_vs_rea_micro} shows that the work by Queiroz et al. had the highest percentages of correct answers regarding micro expressions (all above 40\%, and they did not present videos about Surprise and Disgust). The work by Andreotti et al. had the lowest percentages (all below 20\%, and they did not present a video about Surprise). Concerning macro expressions, as shown in Table~\ref{tab:confusionMatrixEmotions_femaleParticipants_micro}, only Anger and Fear had percentages below 50\%. In Table~\ref{tab:confusionMatrixEmotions_unr_vs_rea_micro}, comparing the three studies, we can see that the work by Andreotti et al. did not have the highest percentage of correct answers when the macro expression was Fear. Our work had the lowest percentages, with Fear and Disgust (in this case, the other two studies did not present a video with Disgust's macro expression) being the exceptions. However, still, in our work, only the macro expressions of Fear and Anger had percentages below 50\%. 

Regarding $H0_1$ %comparison between realistic VH and cartoon VH 
(AVG=29.58\% of realistic VH and AVG=10.74\% of cartoon VH), as well as in Section~\ref{sec:microIso}, we found a significant result (Stats=$3.06$ and \textit{p}=$.01$). \textbf{So we can refute $H0_1$, where people more correctly recognized a micro expression followed by a macro expression in the realistic VH than in the cartoon VH.} While concerning macro expression after micro, the cartoon VH had a higher percentage average (72.91\%) of correct answers than the realistic VH (51.04\%). However, we did not find a significant result, that is, \textbf{we cannot refute $H0_1$ for a macro expression presented after a micro expression.} Regarding $H0_2$, both about a micro expression presented first and a macro expression presented later, we did not find significant results for either case. Even though, for the 2014 VH, the average percentages of correct answers for the micro (AVG=43.10\%) and macro (AVG=67.61\%) expressions were higher than the average percentages presented in our work.  With that, \textbf{we cannot refute $H0_2$ by the fact that, for both micro and macro, people correctly recognized the expressions.} In the comparison between macro and micro expressions of our work, unlike what happened with isolated expressions, we found a significant result (Stats=$3.15$ and \textit{p}=$.01$). Therefore, \textbf{we can say that people answered more correctly about a macro expression after a micro than a micro expression before a macro.}
%SO: tentar verificar se todas as escritas de micro expression estão iguais... umas tem hifen (meu grammarly pediu..) verifica micro e macro
%RU: Corrigido. Mudei todos para micro expression e macro expression.

\begin{table*}[ht]
  \centering
  \caption{
Confusion matrix with the percentages of all participants' responses according to emotional options (Happiness, Sadness, Surprise, Fear, Anger, Disgust) of experiment Part 2 (micro expressions followed by macro expressions). The first seven lines are related to the percentages of the micro expressions, and the last seven to the percentages of the macro expressions.}
  \label{tab:confusionMatrixEmotions_femaleParticipants_micro}
  %\begin{adjustbox}{max width=0.98\linewidth}
  \begin{tabular}{|c|c|c|c|c|c|c|c|c|}
   \hline
   \textbf{Micro} & \textbf{Happiness} & \textbf{Sadness}  & \textbf{Anger} & \textbf{Surprise} & \textbf{Fear} & \textbf{Disgust}& \textbf{No Emotion}& \textbf{I Don't Know}\\
    \hline
    \textbf{Happiness} & \textbf{39.17\%} & 15.00\% & 5.83\% & 18.33\% & 4.17\% & 4.17\% & 3.33\% & 10.00\%\\ 
     \hline
      \textbf{Sadness} & 8.89\% & \textbf{42.22\%} & 4.44\% & 10.00\% & 4.44\% & 10.00\% & 8.89\% & 11.11\%\\ 
      \hline
    \textbf{Anger} & 11.11\% & 14.44\% & \textbf{27.78\%} & 15.56\% & 4.44\% & 10.00\% & 10.00\% & 6.67\%\\ 
     \hline
      \textbf{Surprise} & 8.33\% & 4.17\% & 15.00\% & \textbf{39.17\%}  & 10.00\% & 9.17\% & 7.50\% & 6.67\%\\
      \hline
      \textbf{Fear} & 0.00\% & 3.33\% & 1.67\% & 25.00\% & \textbf{11.67\%} & 15.00\% & 28.33\% & 15.00\%\\
     \hline
      \textbf{Disgust} & 4.17\% & 14.17\% & 7.50\% & 26.67\% & 16.67\% & \textbf{17.50\%} & 6.67\% & 6.67\%\\
      \hline
      \hline
     \textbf{Macro} & \textbf{Happiness} & \textbf{Sadness}  & \textbf{Anger} & \textbf{Surprise} & \textbf{Fear} & \textbf{Disgust}& \textbf{No Emotion}& \textbf{I Don't Know}\\
    \hline
    \textbf{Happiness} & \textbf{62.22\%} & 10.00\% & 6.67\% & 13.33\% & 3.33\% & 4.44\% & 0.00\% & 0.00\%\\ 
     \hline
      \textbf{Sadness} & 10.00\% & \textbf{60.00\%} & 7.78\% & 2.22\% & 3.33\% & 16.67\% & 0.00\% & 0.00\%\\ 
      \hline
    \textbf{Anger} & 16.67\% & 3.33\% & \textbf{36.67\%} & 26.67\% & 6.67\% & 10.00\% & 0.00\% & 0.00\%\\ 
     \hline
      \textbf{Surprise} & 10.00\% & 10.00\% & 0.83\% & \textbf{55.83\%}  & 9.17\% & 14.17\% & 0.00\% & 0.00\%\\
      \hline
      \textbf{Fear} & 5.33\% & 7.33\% & 3.33\% & 32.00\% & \textbf{39.33\%} & 12.67\% & 0.00\% & 0.00\%\\
     \hline
      \textbf{Disgust} & 2.22\% & 11.11\% & 12.22\% & 16.67\% & 5.56\% & \textbf{52.22\%} & 0.00\% & 0.00\%\\
      \hline
%    \bottomrule
  \end{tabular}
  %\end{adjustbox}
\end{table*}

\begin{table*}[ht]
  \centering
  \caption{
Percentages of all options (Happiness, Sadness, Surprise, Fear, Anger, Disgust) of micro isolated animations in relation to the three studies (Our, Andreotti et al.~\cite{andreotti2021perception}, and Queiroz et al.~\cite{queiroz2014investigating}).}
  \label{tab:confusionMatrixEmotions_unr_vs_rea_macro}
  %\begin{adjustbox}{max width=0.98\linewidth}
  \begin{tabular}{|c|c|c|c|c|c|c|}
   \hline
   \textbf{Micro Isolated} & \textbf{Happiness} & \textbf{Sadness}  & \textbf{Anger} & \textbf{Surprise} & \textbf{Fear} & \textbf{Disgust}\\
    \hline
    \textbf{Our - Realistic} & {83.33\%} & \textbf{80.00\%}  & {50.00\%} & {90.00\%} & {36.67\%} & \textbf{86.67\%}\\ 
    \hline
    \textbf{Andreotti et al. \cite{andreotti2021perception} - Unrealistic} & {43.20\%} & {7.40\%}  & {61.70\%} & {69.10\%} & {16.00\%} & {30.90\%}\\ 
    \hline
    \textbf{Queiroz et al. \cite{queiroz2014investigating} - Realistic} & \textbf{87.50\%} & {75.60\%}  & \textbf{92.26\%} & \textbf{92.86\%} & \textbf{45.24\%} & {75.60\%}\\ 
    \hline
  \end{tabular}
  %\end{adjustbox}
\end{table*}

\begin{table*}[ht]
  \centering
  \caption{
Percentages of all options (Happiness, Sadness, Surprise, Fear, Anger, Disgust) of micro animations when followed by macro animations in relation to the three studies (Our, Andreotti et al.~\cite{andreotti2021perception}, and Queiroz et al.~\cite{queiroz2014investigating}). The first four lines are related to the percentages of the micro expressions, and the last four to the percentages of the macro expressions.}
  \label{tab:confusionMatrixEmotions_unr_vs_rea_micro}
  %\begin{adjustbox}{max width=0.98\linewidth}
  \begin{tabular}{|c|c|c|c|c|c|c|c|c|}
   \hline
   \textbf{Micro} & \textbf{Happiness} & \textbf{Sadness}  & \textbf{Anger} & \textbf{Surprise} & \textbf{Fear} & \textbf{Disgust}\\
    \hline
    \textbf{Our - Realistic} & {39.17\%} & {42.22\%}  & {27.78\%} & \textbf{39.17\%} & {11.67\%} & \textbf{17.50\%}\\ 
    \hline
    \textbf{Andreotti et al. \cite{andreotti2021perception} - Unrealistic} & {13.33\%} & {2.50\%}  & {16.66\%} & {-} & {10.00\%} & {11.25\%}\\ 
    \hline
    \textbf{Queiroz et al. \cite{queiroz2014investigating} - Realistic} & \textbf{56.75\%} & \textbf{47.02\%}  & \textbf{48.41\%} & \textbf{-} & \textbf{20.24} & \textbf{-}\\ 
    \hline
    \hline
   \textbf{Macro} & \textbf{Happiness} & \textbf{Sadness}  & \textbf{Anger} & \textbf{Surprise} & \textbf{Fear} & \textbf{Disgust}\\
    \hline
    \textbf{Our - Realistic} & {62.22\%} & {60.00\%}  & {36.67\%} & {55.83\%} & \textbf{39.33\%} & \textbf{52.22\%}\\ 
    \hline
    \textbf{Andreotti et al. \cite{andreotti2021perception} - Unrealistic} & \textbf{88.13\%} & \textbf{72.50\%}  & \textbf{85.00\%} & \textbf{90.80\%} & {28.12\%} & \textbf{-}\\
    \hline
    \textbf{Queiroz et al. \cite{queiroz2014investigating} - Realistic} & {86.31\%} & {72.02\%} & {70.24\%} & {83.33\%} & {26.19\%} & {-}\\ 
     \hline
%    \bottomrule
  \end{tabular}
  %\end{adjustbox}
\end{table*}

\section{Discussion}
\label{sec:discussion}

This section aims to discuss the results regarding the hypotheses of our work. Regarding $H0_1$ (''There is no significant difference in the perception of micro and macro facial expressions between realistic and cartoon virtual humans"), talking about micro expressions presented in separate videos, as happened between the studies by Queiroz et al.~\cite{queiroz2014investigating} (realistic) and Andreotti et al.~\cite{andreotti2021perception} (cartoon), people in our work (with realistic characters) more easily recognized micro expressions than people in Andreotti's work (cartoon). Andreotti et al. raised two hypotheses with their results on micro expressions. \textit{i)} First, the evolution of technology. The authors claim that the work developed by Queiroz et al. was done in 2014 and since then, technology has advanced a lot over time, so with more realistic animations concerning the face, micro expressions may have become more imperceptible. However, our results showed that no matter whether the technology is old or new, the result was the same. And \textit{ii)} The type of realism of the character (while Andreotti et al. used a cartoon character, Queiroz et al. used a realistic character) may have influenced it. This hypothesis seems to be what happened with our results as well. Regarding the isolated macro expression, as the work by Andreotti et al. only presented the mean percentage value, we could not compare using statistical methods. However, looking at the average percentages, we can see that the participants of the two studies had percentages close to 80\%, that is, both in relation to a cartoon VH and a realistic VH, the isolated macro expression tends to be recognized. Still regarding $H0_1$, but talking about micro expressions presented before macro expressions, again, people recognized the micro expression of the realistic VH (both, in our work and in Queiroz et al.) more than the cartoon. Regarding the macro expression presented after a micro, statistically speaking, people recognized expressions from realistic and cartoon VHs similarly.

Concerning $H0_2$ (''There is no significant difference in the perception of micro and macro facial expressions between realistic virtual humans in 2014 and those in 2023"), the results did not show significant changes in perception between the realistic VH from 2014 (work by Queiroz et al. \cite{queiroz2014investigating}, AVG=78.17\%) and the realistic VH from 2023 (our work, AVG=71.11\%). This aligns with our hypothesis $H0_2$ that cannot be refuted, indicating that although advancements in technology and animation have occurred over the nine-year gap, the accuracy of emotion recognition in realistic characters remained relatively consistent. This might imply that certain aspects of human emotion perception are not as influenced by technological advancements, and other factors like human psychology and cognition may significantly impact emotion understanding. In addition, we had a doubt that can be explored in a possible future work. First, this result shows us that people from each of the two studies (ours and Queiroz et al.) evaluated the realistic VHs of their times. However, what if people in 2023 (or later) evaluated micro and macro expressions created with old technologies? In the work by Araujo et al.~\cite{araujo2021perceived}, the authors evaluated the comfort perceived by people from 2020 concerning VHs created with older technologies and compared the comfort perceived by people from 2012 with these same VHs. The results showed that both people in 2012 and 2020 felt comfortable similarly to old characters.
%SO: relative a personagens passados não eram similares?
%Vic: vdd.. mudei
Bringing this result to our work, would the same happen in relation to micro and macro expressions? % Gi: ainda vou botar uma referencia para esse paragrafo for mantido, já que parece que descrevi muito brevemente %Vic: gostei! Vic: acrescentei umas coisinhas

Additionally, our study reveals that participants consistently recognized macro expressions more accurately than micro expressions, even when both were displayed together. It is worth pointing that macro expressions are more visible and easier to detect as they involve more significant facial movements and expressions compared to subtle micro expressions. This finding corroborates with previous studies \cite{queiroz2014investigating} where macro expressions had higher recognition rates than micro expressions. This is most evident when a macro expression is presented after a micro expression, where the perception values significantly decrease. As pointed out in the discussion of Andreotti et al.'s work, when people saw both micro and macro expressions in isolation, the difference in expression recognition was smaller than when people saw a micro followed by a macro expression. This makes sense, as micro expressions are expressions that happen quickly compared to macro expressions. %So, if the designer, programmer, or any professional wants a micro expression to be recognized, the ideal is not to present it accompanied by a macro expression. 
%SO: estou em dúvida sobre isso acima.... micro expressions servem apra não serem notadas :)
%Vic: verdade, podemos tirar isso.. mas ai nesse caso o foco é se o profissional QUER q alguem perceba.. mas eu entendi oq a senhora falou
%Vic: comentei a frase e acrescentei outra

%Gi: se o victor gostar continuarei depois da festinha
%Vic: kkkkk pode ir %Vic: gostei tb! so precisa acrescentar que fizemos duas comparações micro vs macro.. uma em relacao a isoladas e a outra qndo era micro seguida de macro (essa q deu diferença significativa).. Ai podemos falar que, a importancia de mostrar elas separadamente para que as pessoas reconheçam, ja que uma seguida da outra atrapalha, e acaba q a macro fica mais evidente Vic: acrescentei umas coisinhas

%\item $H0_1$ - There is no significant difference in the perception of micro and macro facial expressions between realistic and cartoon virtual humans;
%\item $H0_2$ - There is no significant difference in the perception of micro and macro facial expressions between realistic virtual humans in 2014 and those in 2023;

\section{Final Considerations}
\label{sec:final_considearions}

In the present work, we conducted a perceptive study on recognizing micro and macro expressions in realistic VHs. We re-visited two studies from the literature and compared them with our results.  These studies were proposed by Andreotti et al.~\cite{andreotti2021perception}, which used a cartoon VH, and Queiroz et al.~\cite{queiroz2014investigating}, that used a realistic VH, and intend to answer our two research hypotheses.

We partially refuted $H0_1$, showing that people tend to better recognize micro facial expressions in realistic VHs than cartoons and similarly recognize macro expressions. We did not refute $H0_2$, showing that people tend to similarly recognize micro and macro realistic VH expressions from different years. In possible future work, we intend to include analyses on gender since there are scientific studies (\cite{bailey2017gender, nag2020gender, zibrek2013evaluating, zibrek2015exploring, durupinar2022facial}) that show that perception of emotions is different depending on the gender of VH and participant. Furthermore, we intend to use old and new technologies to create realistic VHs and evaluate whether the recognition of micro and macro expressions remains similar. In addition, we also plan to use a virtual human created by a designer to include in future analyses.

\begin{acks}
We would like to thank CNPq and CAPES for partially supporting this work.
\end{acks}

\bibliographystyle{ACM-Reference-Format}
\bibliography{sample-base}

\end{document}